\documentclass[conference]{IEEEtran}
\IEEEoverridecommandlockouts
\usepackage{cite}
\usepackage{amsmath,amssymb,amsfonts}
\usepackage{algorithmic}
\usepackage{graphicx}
\usepackage{textcomp}
\usepackage{xcolor}
\def\BibTeX{{\rm B\kern-.05em{\sc i\kern-.025em b}\kern-.08em
    T\kern-.1667em\lower.7ex\hbox{E}\kern-.125emX}}
\begin{document}

\title{(M)SLA$e$-Net: Multi-Scale Multi-Level Attention embedded Network for Retinal Vessel Segmentation}

%\author{ Paper ID: 
%}

\author{\IEEEauthorblockN{1\textsuperscript{st} Shreshth Saini}
\IEEEauthorblockA{\textit{Department of Electrical Engineering} \\
\textit{Indian Institute of Technology Jodhpur}\\
India\\
saini.2@iitj.ac.in}
\and
\IEEEauthorblockN{2\textsuperscript{nd} Geetika Agrawal}
\IEEEauthorblockA{\textit{Department of Electrical Engineering} \\
\textit{Indian Institute of Technology Jodhpur}\\
India\\
agrawal.6@iitj.ac.in}
\and
}

\maketitle

\begin{abstract}
Segmentation plays a crucial role in diagnosis. Studying the retinal vasculatures from fundus images help identify early signs of many crucial illnesses such as diabetic retinopathy. %, retinal vascular diseases, cardiac diseases, and ophthalmologic diseases. %Given the challenging nature of the task of segmentation of retinal vessels due to poor contrast and the presence of noise and artefacts. Until now many convolutional neural networks (CNNs) based methods have set-up a benchmark on various datasets but there’s a lot of room for improvements, especially when it comes to precisely segmenting the small, thin, obscured, and faded vessel.
\textcolor{black}{Due to the varying shape, size, and patterns of retinal vessels, along with artefacts and noises in fundus images, no one-stage method can accurately segment retinal vessels.} In this work, we propose a multi-scale, multi-level attention embedded CNN architecture ((M)SLAe-Net) to address the issue of multi-stage processing for robust and precise segmentation of retinal vessels. \textcolor{black}{We do this by extracting features at multiple scales and multiple levels of the network, enabling our model to holistically extracts the local and global features. Multi-scale features are extracted using our novel dynamic dilated pyramid pooling ($D$-DPP) module. We also aggregate the features from all the network levels. These effectively resolved the issues of varying shape and artefacts and hence the need for multiple stages.} To assist in better pixel level classification, we use the Squeeze and Attention ($SA$) module, a smartly adapted version of the Squeeze and Excitation ($SE$) module for segmentation tasks in our network to facilitate pixel-group attention. Our unique network design and novel $D$-DPP module with efficient task specific loss function for thin vessels enabled our model for better cross data performance. Exhaustive experimental results on DRIVE, STARE, HRF, and CHASE-DB1 show the superiority of our method.% in obtaining more robust, reliable, and precise masks over existing state-of-the-art methods. 

\end{abstract}


\section{Introduction}
\label{sec:intro}
Segmentation of retinal vasculatures plays a vital role in the diagnosis of retinal diseases and many other systemic diseases \cite{1}
, such as cardiovascular, diabetic retinopathy%\cite{3}, 
, and hypertension. %\cite{4}. 
Further, retinal vessels are useful in blow flow analysis and biometric recognition. %\cite{5}. 
There is an urgent need for intelligent automated methods for retinal vessel segmentation as currently, the task is being done manually by expert ophthalmologists, which is an extremely error-prone, and time-consuming process lacking reproducibility. %The largely varying shape, size, and pattern, thin and faded vessels in noisy images which may present irregular illumination, sensor noise, blur, poor contrast along with the unavoidable presence of anatomical structures and artefacts such as foveas, macula, and hemorrhages, cotton wool spots, and exudates make the segmentation task even more challenging. 
%Almost all techniques classical or deep learning based fail to perfectly segment the complete retinal vessels, some perform poor with simultaneous segmentation of the major and thin vessels while some fail to look beyond noise and abnormalities of fundus images. Another bottleneck for deep learning based methods is the lack of availability of large annotated retinal vessel segmentation dataset, the largest number of annotated fundus images in publicly available datasets such as DRIVE[6], STARE[7], CHASE-DB1[8], and HRF[9] has only 20 images for training purpose.

\begin{figure}
    \centering\includegraphics[width=1\linewidth]{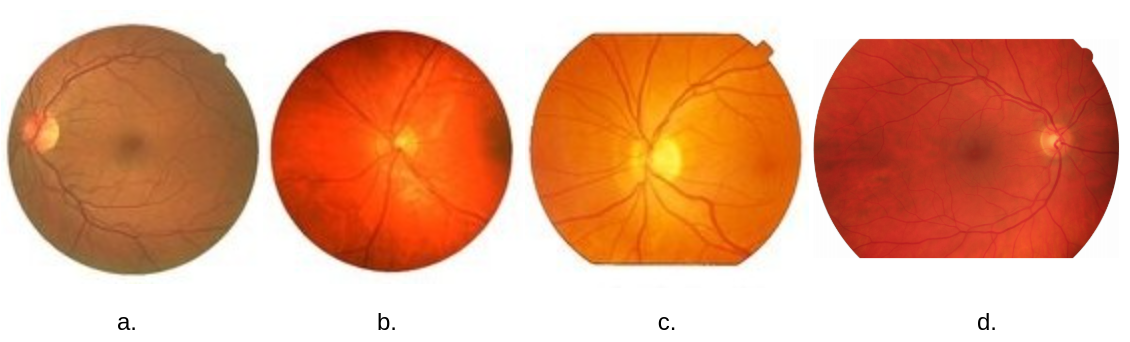}
    \caption{Fundus Images. a. DRIVE, b. STARE, c. CHASE-DB1, d. HRF dataset.}
    \label{fig:dataset}
\end{figure}

Classical approaches focused on filters to extract vessels' boundaries followed by some post-processing steps for final segmentation maps. %\cite{10}. %11 These approaches involved the edge detection process, morphological operations, vessel tracking methods, and other image processing techniques which inherently lacked generalisability. 
Fundus images from different datasets have essentially different appearances (shown in Fig.\ref{fig:dataset}), and hence classical approaches tend to perform poorly compared to benchmarks. On the other hand, learning based methods \cite{3}
outperform classical methods due to better generalisability and relevant and complex feature extraction. With recent advances in deep learning where CNN based methods have shown a tremendous superiority in performance, %keeping up with the state-of-the-art and trend, 
almost all researchers have dwelled deep into the deep learning based methods for retinal vessel segmentation.

U-Net \cite{16}, which has become the standard framework for many medical image segmentation tasks mainly due to its unique network and skip connections, many networks have used a similar design \cite{saini2, saini3}. U-Net \cite{16} struggles to segment when deployed for retinal vessel segmentation. Other CNN designs derived from U-Net \cite{16} face somewhat similar issues of missing thin and faded vessels and segmenting optic disks \cite{17, 8}. Recently more and more deep learning based methods are focusing on developing network modules to extract the relevant features \cite{24,saini1}.   
At the same time, researchers with novel contributions, custom convolution, loss function, and training strategy have shown some promising results. Alom et al. \cite{20} propose the R2-UNet, which incorporates the efficient residual blocks and recurrent convolutional layer into the UNet architecture with patch based segmentation lacking the global context. In \cite{21}, the authors propose an end-to-end dense dilated CNN model, the output of which is combined with probability regularized walk for vessel segmentation. Fan et al. \cite{22}, developed a novel octave convolution based architecture for efficient multi-frequency feature extraction for accurate vessel segmentation but was unable to address the artefact obstructions. Wang et al. \cite{23}, in their work CTF-Net, propose the Coarse-to-fine deep network to tackle low contrast and noise issues in retinal vessel segmentation. \textcolor{black}{To the best of our knowledge, no single method can efficiently address all the challenges.}
%Many attention based modules have been adapted for retinal vessel segmentation tasks \cite{25, 26}, recently proposed SA-Net \cite{24} for natural image segmentation adopts a pixel-group attention strategy in their work. To the best of our knowledge the Squeeze and attention module has not been incorporated in retinal vessel segmentation tasks whereas the earlier version of it namely squeeze and excitation, spatial and channel attention modules has been successfully used in the medical segmentation domain. 

\begin{figure}
    \centering\includegraphics[width=1\linewidth, height=1.1\linewidth]{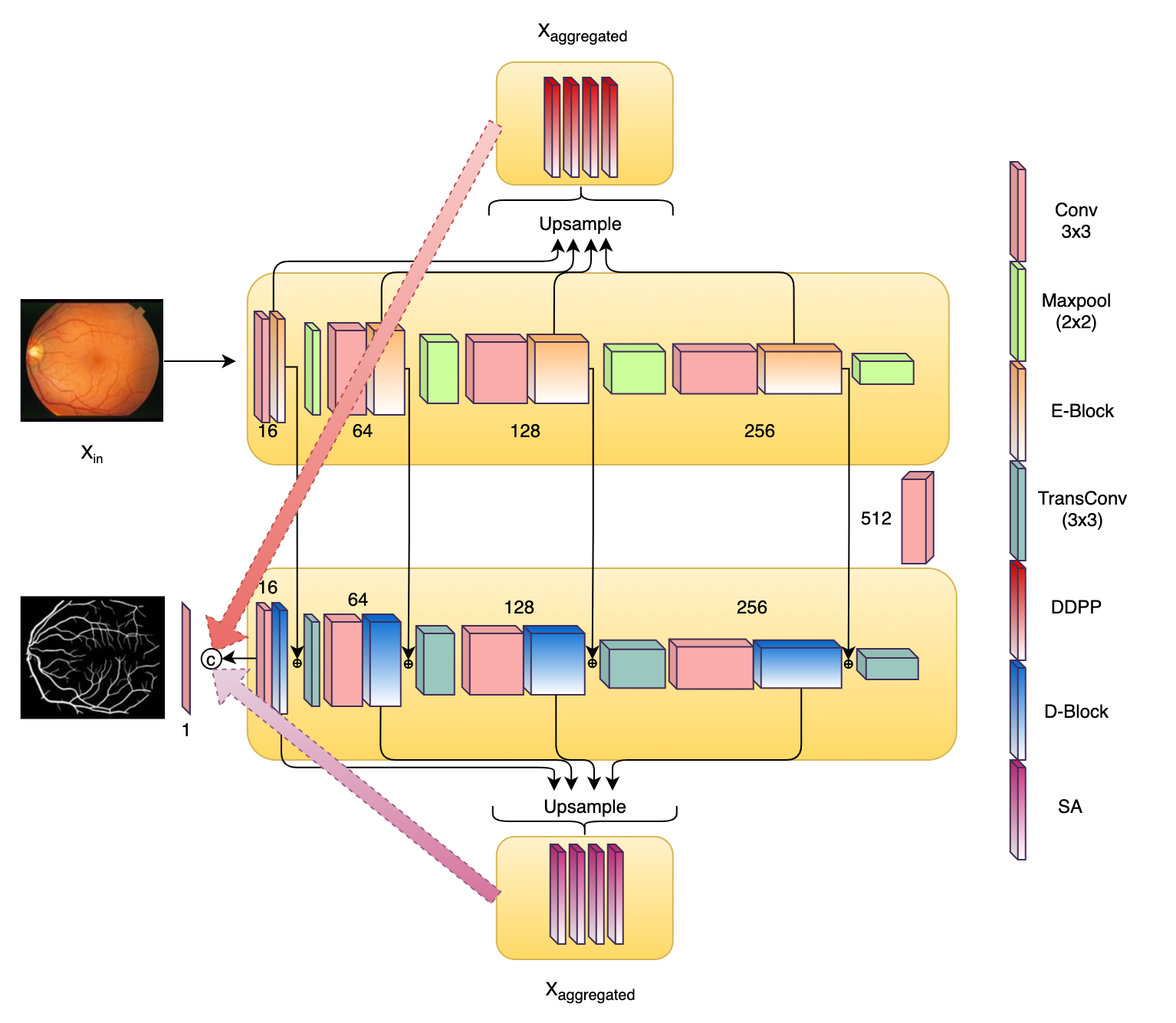}
    \caption{Overview of our proposed Network (M)SLAe-Net. Aggregated feature maps from encoder and decoder are combined with decoder output.}
    \label{fig:network}
\end{figure}

\textcolor{black}{In this work, we propose an end-to-end efficient and compact Multi-Scale, Multi-Level Attention embedded Network ((M)SLAe-Net) in order to address the major challenges, i.e. varying shape, size and artefacts. Our network consists of two parts overall, encoder and decoder. Each block in the encoder ($E-Block$) is a combination of our novel dynamic dilated pyramid pooling ($D$-DPP) and squeeze and attention ($SA$) modules. These modules equip the encoder with attention and multi-scale feature extraction abilities; which enabled the model to look beyond the obstruction and noises in fundus images. Further details are given in section \ref{sec:methods}.} In the decoder part, each block ($D-Block$) has only $SA$ modules in them. There are skip connections present from the encoder to the decoder for efficient gradient flow. At each level of the encoder, we extract consistent multi-scale features, which are later fused with highly condensed attention based features extracted from all the levels of the decoder as shown in Fig \ref{fig:network}, \textcolor{black}{virtually giving our network a multi-stage design. This unique network design improved our segmentation maps adding robustness to them.}

\section{Method}
\label{sec:methods}

In this section, we discuss our proposed (M)SLAe-Net and its modules. A detailed discussion on blocks used in the encoder as well as the decoder parts is provided. The overall network is shown in Fig. \ref{fig:network}. %Finally, We describe the complete end-to-end network design and its features for retinal vessel segmentation. As depicted in Fig. 2, we limit the network depth to 4 levels and have skip connection from encoder layers to corresponding decoder layers, it also gathers the multi scale features from all the levels details of which are also discussed in this section.
  
\begin{figure}
    \centering\includegraphics[width=1\linewidth, height=1\linewidth]{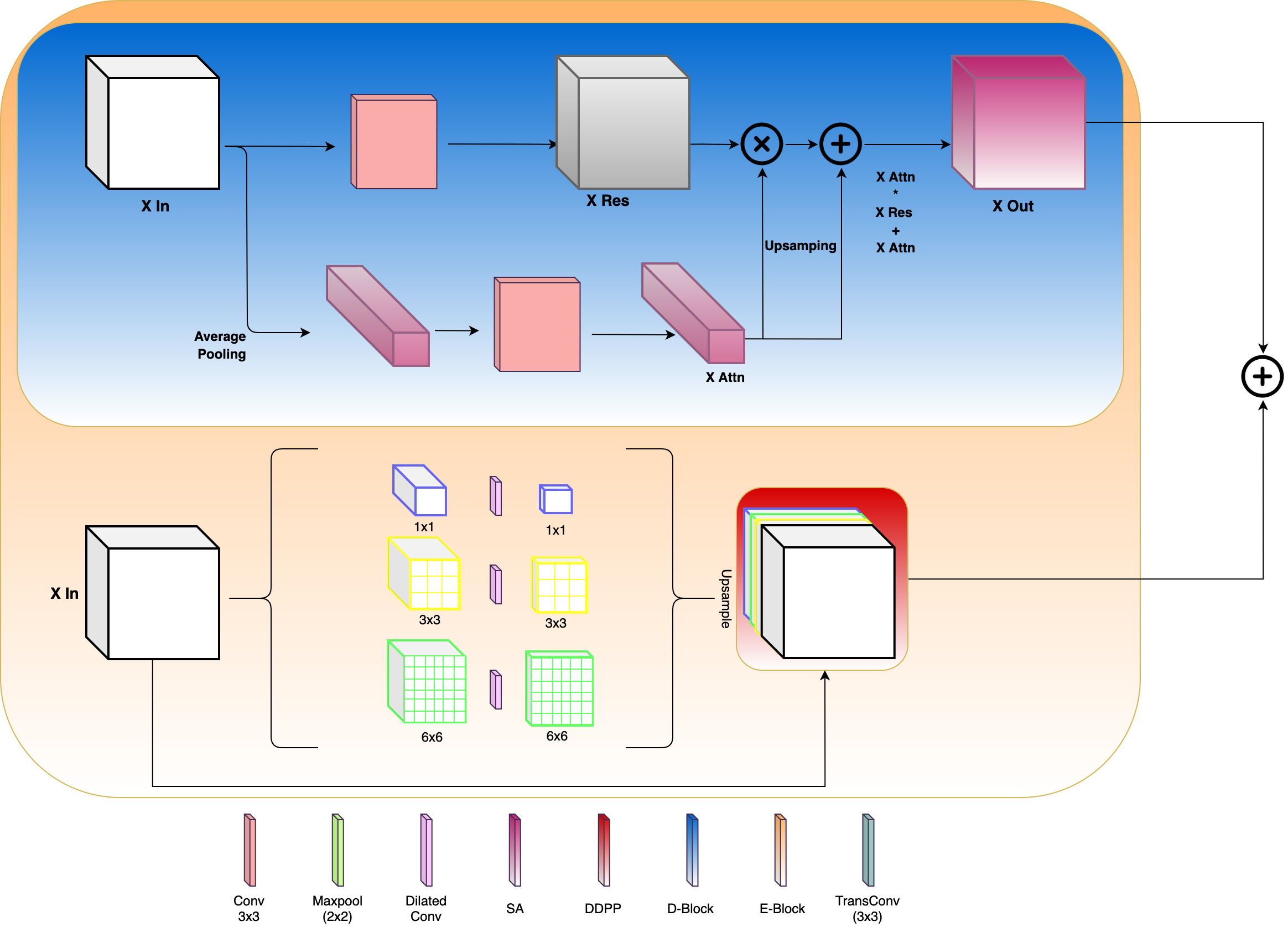}
    \caption{Network modules. $E-Block$, $D-Block$, $D$-DPP module, and $SA$ module. In decoder, only $SA$ modules are used, skip connections are extracted from $D$-DPP modules in encoder.}
    \label{fig:modules}
\end{figure}

%\subsection{Encoder} 
%\label{sec:encoder}
\textbf{Encoder: }%For the segmentation task most widely used network structure is auto-encoder which consist of an encoder and a decoder.%which downsamples the input followed by a decoder taking the downsampled features map as input to generate a mask. 
U-Net \cite{16} is essentially an auto-encoder with skip connections from the encoder layers to the corresponding decoder layers to allow the flow of global information and gradients across the network. In our encoder, we use $E-Block$ rather than a simple convolutional layer, which consists of the $D$-DPP and $SA$ modules. When an input feature map ($X_{in}$) enters the $E-Block$, it gets shared over the two modules. %SA and DDPP modules each output a feature map which are combined together to output the final feature map from E-Block.
Output feature map has extracted features at multiple scales from the $D$-DPP module and local context and pixel-group attention from the $SA$ module. The skip connections are taken from the output of $E-Block$, whereas for aggregation of multi-level features, the outputs from $D$-DPP blocks are taken as depicted in Fig. \ref{fig:modules}. \textcolor{black}{ Multi-scale feature processing at each encoder level allows the network to capture relevant features and look beyond obstructions.}

\begin{align}
    \label{eq:encoderrr}
    \chi_{bottleneck} = \mathcal{E}_{encoder}(X_{in}) \\ 
    \label{eq:aggregated1}
    \chi_{aggregated} = \sum_{i}^{4} (D_{DPP}(\chi_{i}) + SA(\chi_{i}))
\end{align}

In Equation \ref{eq:encoderrr}, $\mathcal{E}_{encoder}$ presents the complete encoder part, $\chi_{bottleneck}$ is the output feature map from the encoder. In Equation \ref{eq:aggregated1}, $\chi_{aggregated}$ is aggregated feature map from $SA$ modules and $D$-DPP modules. $\chi_{1}$ is convoluted input image $X_{in}$. Equation \ref{eq:DDPP} shows the working of our $D$-DPP module.

%\subsection{DDPP Module} 
%\label{sec:ddpp}
\textbf{$D$-DPP Module: }In our dynamic dilated pyramid pooling module, we present the use of dilated convolution dynamically varying with the network level. If it is to be used at the input level, then the dilation rate is kept at 1, the rate is 2, 3, and 4 at levels 2, 3, and 4, respectively. Within a pyramid, the dilation rate is kept the same, ensuring that features are appropriately extracted with relevant receptive fields. For pyramid pooling, we propose a 3 level pooling 1x1, 3x3, and 6x6. The output from each scale is upsampled accordingly to give back the feature size the same as that of the input feature, which combined with dilated convolutional, captures the essentially larger vessels and patterns, reducing the discontinued vessel masks. Equation \ref{eq:DDPP} and Fig. \ref{fig:modules} depict our $D$-DPP module. 
%Output from this module is combined with output from the SA module and is also sent for aggregation from all levels which is discussed in the (M)SLAe-net section. 

\begin{align}
     \label{eq:DDPP}
     \chi_{out} = (\sum_{i=1,3,6} \Upsilon(C_{dilated}(Kernel_{iXi}(\chi_{in})))) + \chi_{in}
\end{align}

In Equation \ref{eq:DDPP}, $C_{dilated}$ is dilated convolution, $\Upsilon$ is upsampling step. 

%\subsection{SA Module} 
%\label{sec:sa}	
\textbf{$SA$ Module: }Many attention based convolutional blocks have been used widely in the deep learning domain for more reliability and explainability. While the channel and spatial attention module select the most relevant spatial regions or channels from a feature map, they can not often be incorporated for segmentation tasks due to the poor handling of feature dependencies over the spatial regions and across channels. Squeeze and Attention \cite{24} module tackle this issue by considering all inter-dependencies.  It brings the pixel-group attention by proposing attention on convolutional channels. SA module extracts and focuses on local context giving more confident and robust vessel segmentation even for thin vessels. Equation \ref{main_sa} shows the actual operation of the $SA$ module. Fig. \ref{fig:modules} shows the $SA$ module. {Let convolutional block be $C_{block}$, then : }

\begin{align}
    \label{main_sa}
    \chi_{out} = \chi_{attn} * \chi_{res} + \chi_{attn} \\
    \chi_{attn} = \Upsilon_{SA}(\sigma (C_{block}(\rho(\chi_{in}))))
\end{align}

{Input feature map $\chi_{in}$ is passed through a $C_{block}(\cdot)$ to get $\chi_{res}$. This residual input is weighted with $\chi_{attn}$, $\Upsilon_{SA}$ upsamples the processed attention map for direct multiplication and addition with the $\chi_{res}$. Here, $\rho$ and $\sigma$ represents average pooling operation and sigmoid activation function respectively.}

%\subsection{Decoder}
%\label{sec:decoder}	
\textbf{Decoder: }We use $D-Blocks$ in the decoder (see Fig. \ref{fig:modules}), which has only the $SA$ module in it. Skip connections from the encoder part allow the decoder to produce location precise segmentation masks. We extract features that are aggregated from each $SA$ module to be combined with aggregated features from the encoder part. Aggregated features from the encoder being extracted with $D$-DPP have global contextual information, and aggregated features from the decoder have condensed local contextual information as those features were essentially processed through cascaded $SA$ modules. A combined feature map is added to the output of the decoder to give the final output. 

%\subsection{(M)SLAe-Net}
%\label{sec:mslae-net}
\textbf{(M)SLAe-Net: }Leveraging the multi-scaled feature extraction at every depth level of the network with embedded attention in every feature map enables our (M)SLAe-Net to predict precise and robust retinal vessel segmentation. With the aggregation of features from all the levels, we ensured that our final output has masks for complete vessels, which is the bottleneck for the most state-of-the-art methods. Our model consists of a 3x3 convolutional layer with ReLu activation function followed by Batch Normalisation. The number of channels are 16, 64, 128, 256 in the encoder and 512 at the bottleneck. The decoder has the same channels as that of the encoder. 

\section{Experimentation and Results}
\label{sec:results}
This section provides comprehensive experimental details and performance comparison for our proposed (M)SLAe-Net. %We also try to discuss the ablation study for our work in this section.

%\subsection{Dataset Description} 

\textbf{Dataset Description: }We used four retinal vessel segmentation datasets for experimentation purpose. DRIVE, CHASE-DB1, HRF, and STARE each dataset is  publicly available for research purposes. DRIVE, CHASE-DB1, HRF, and STARE have image resolutions of 565x584, 999x960, 3304x2336, and 700x605 respectively. %Largest dataset (HRF) have 45 images often which only 15 healthy patient's images are employed for segmentation task. 
%which stands for Digital Retinal Images for Vessel Extraction have 40 color fundus images each with a resolution of 565x584. %It has 20 images for training purposes and rest for testing. CHASE-DB1[8] which is provided by British Children’s Hearing and Health Research Project contains a total of 28 fundus images with a resolution of 999x960. %In this dataset, 8 images are kept aside for testing purposes only. 
%HRF[9] (High Resolution Fundus Image Database) released in 2011 has the largest 45 fundus images whereas the distribution states of total, 15 images for each healthy, glaucoma, and diabetic class are present. Structural Analysis of the Retina (STARE[7]) dataset holds a total of 30 fundus images having a resolution of 700x605 where 10 images are abnormal and only 20 can be used. 
\textcolor{black}{We train our model once with DRIVE and test on all. }
%\textbf{For training and testing purposes, in the absence of a known split, we employed a 5-fold cross validation strategy. All the images were resized to 512x512 and normalised before feeding into the model.} %No other pre-processing step was incorporated and images were used in raw form. 
We utilised the online data augmentation method with only horizontal and vertical flip, and horizontal and vertical shift.
 
%\subsection{Training Setup, Results and Evaluations }
\textbf{Training Setup, Results and Evaluations: }We have utilised the Tensorflow library for the complete implementation of our method. We performed experiments on Nvidia V100 GPU with a memory of 32 GBs. %We limited the use of GPUs to 1 only. 
We kept a batch size of 22 and optimised our model using Adam optimiser with a learning rate of 1e-4. The model was trained for about 70 epochs from scratch with all the weights initialised with the standard He-Normal distribution.

For the \textbf{loss function}, rather than choosing standard dice loss or binary cross-entropy loss, we use a task specific elastic interaction-based loss \cite{36} ($EI-loss$). $EI-loss$ was introduced for retinal vessel segmentation tasks keeping in mind that vessels are continuous and consistent structures to increase the overall performance. We kept the value of hyperparameter $\alpha$ = 0.50, and $\beta$ = 0.25 in the Hardtanh (smoothing Heaviside) function in $EL-loss$. Please refer to the paper for more details.

\begin{figure}
    \centering\includegraphics[width=\linewidth]{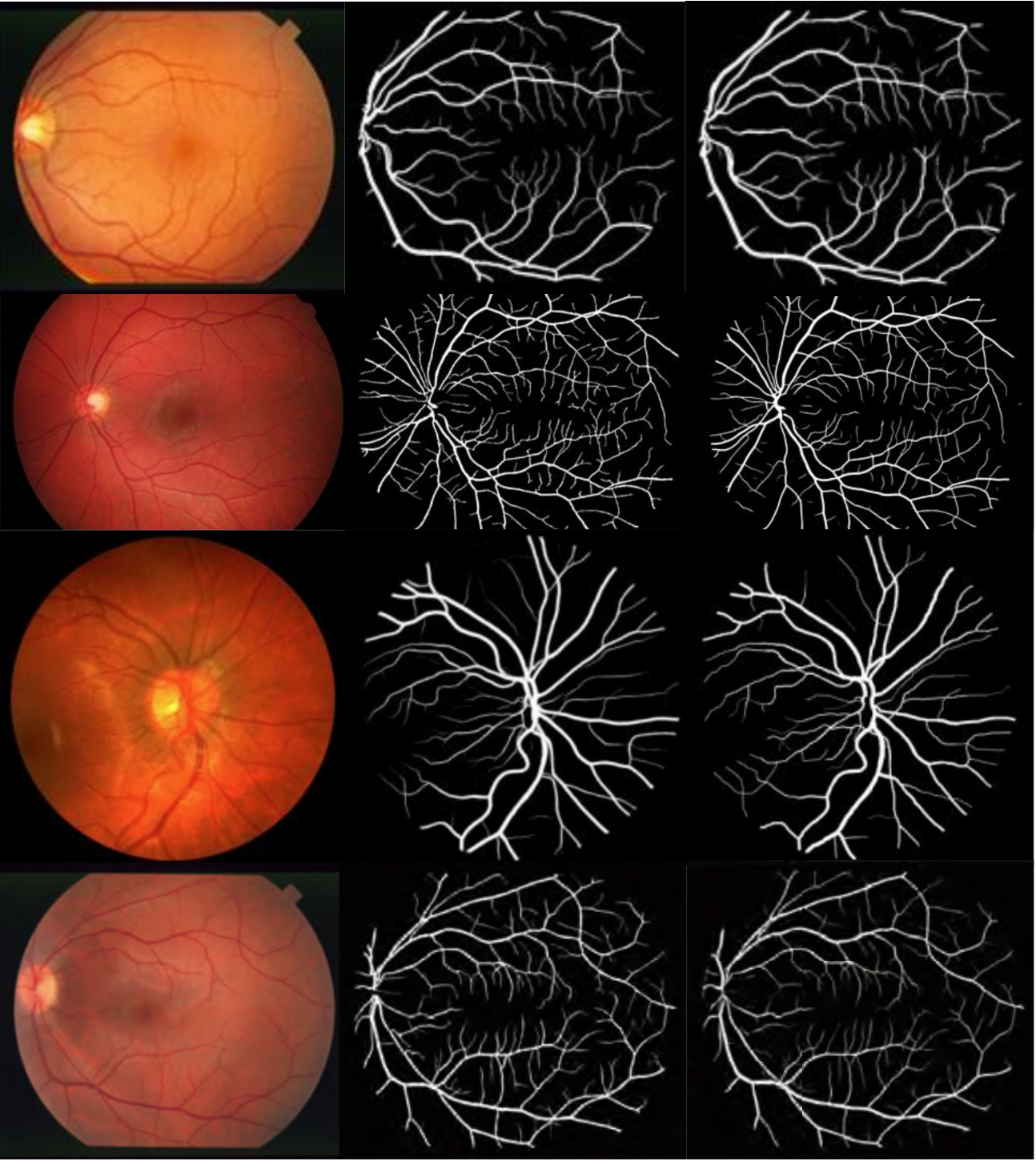}
    \caption{Qualitative Results. From top to bottom row shows results on DRIVE, HRF, CHASE, and STARE. Left most is input fundus image, middle is ground truth mask, and right most is predicted mask.}
    \label{fig:results}
\end{figure}

For the \textbf{evaluation purpose}, we compare our model on metrics accepted across the benchmark for retinal vessel segmentation. Namely, Specificity (Sp), Sensitivity (Se), Accuracy (Acc), and Area Under the Receiver Operating Characteristic Curve (AUROC). 

\begin{align}
    \label{eq:metrics1}
    Sp = \frac{(\Bar{Y} * \Bar{Y'})}{((\Bar{Y} * \Bar{Y'}) + (Y' * \Bar{Y}))} \\
    Se = \frac{(Y * Y')}{((Y * Y') + (\Bar{Y'} * Y))}
\end{align}

\begin{align}
    \label{eq:metrics2}
    Acc = \frac{((Y * Y') + (\Bar{Y} * \Bar{Y'}))}{((Y * Y') + (Y' * \Bar{Y}) + (\Bar{Y} * \Bar{Y'}) + (\Bar{Y'} * Y))} 
\end{align}

In above equations $Y$, $Y'$, $\Bar{Y}$, $\Bar{Y'}$, are foreground ground truth, foreground prediction, background ground truth, and background predictions respectively.  

Table \ref{tab:final_result} shows the comparative analysis on different datasets. It can be observed that our method outperforms on Sensitivity, Accuracy, and AUROC with an average margin of ~0.5\%-1.0\%. Fig. \ref{fig:results} shows the qualitative results, and it can be seen that even the thin and faded vessels were precisely segmented in noisy and uneven illuminated fundus images. \textcolor{black}{Our method achieved the results in an end-to-end manner without the need for multi stage processing. Multi-level feature aggregation boosted our overall performance through the better gradient and information flow. Our unique $D$-DPP module extracted information at multiple scales giving our network the ability to capture varying vessel patterns.} In table \ref{tab:final_result}, the three rows show the results on HRF \cite{9} for the ablation study. We experimented with the absence of the $D$-DPP module and the $SA$ module (replaced by a simple convolutional block). As depicted in the results, $D$-DPP plays a vital role in performance improvement, and so does the $SA$ module. \textcolor{black}{With the $D$-DPP module being absent, the performance was significantly reduced to 0.762, 0.935, and 0.949 for sensitivity, specificity, and accuracy, respectively, compared to (M)SLAe-Net, which are lesser than most of the state-of-the-art methods; this was mainly because the network could not capture thin and terminating vessels; the same was confirmed qualitatively. A similar pattern is observable for the $SA$ module, where vessels were not captured in regions with obstructions. In the final model, where a combination of both the modules with multi-scale feature aggregation collectively improves the performance by efficiently extracting and gathering features. With our modules and network design precisely targeting the challenges related to fundus images, we were able to produce more robust and precise vessel segmentation. }

\begin{table}[]
\centering
%\small 
\caption{Comparative Analysis with state-of-the-art and Ablation study results.}
\label{tab:final_result}
\scalebox{1}{
\begin{tabular}{ccccc}
\hline
Method                  & Se              & Spe             & Acc             & AUROC           \\ \hline
\multicolumn{5}{c}{DRIVE}                                                                       \\ \hline
MS-NFN\cite{27}         & 0.7844          & 0.9819          & 0.9567          & 0.9807          \\
DUNet\cite{17}          & 0.7963          & 0.9800          & 0.9566          & 0.9802          \\
CAR-UNet\cite{28}       & 0.8135          & \textbf{0.9849} & 0.9699          & 0.9852          \\
RSAN\cite{29}                    & 0.8149          & 0.9839          & 0.9691          & 0.9855          \\
(M)SLAe-Net (Our)       & \textbf{0.8189} & 0.9821          & \textbf{0.9705} & \textbf{0.9870} \\ \hline
\multicolumn{5}{c}{CHASE}                                                                       \\ \hline
MS-NFN\cite{27}         & 0.7538          & \textbf{0.9847} & 0.9637          & 0.9825          \\
DUNet\cite{17}          & 0.8155          & 0.9752          & 0.9610          & 0.9804          \\
CAR-UNet\cite{28}       & 0.8439          & 0.9839          & 0.9751          & \textbf{0.9898} \\
RSAN\cite{29}                    & 0.8486          & 0.9836          & 0.9751          & 0.9894          \\
(M)SLAe-Net (Our)       & \textbf{0.8513} & 0.9810          & \textbf{0.9791} & 0.9896          \\ \hline
\multicolumn{5}{c}{SATRE}                                                                       \\ \hline
R2U-Net\cite{30}        & 0.7756          & 0.9820 & 0.9634          & 0.9815          \\
DUNet\cite{17}          & 0.7595          & \textbf{0.9878} & 0.9641          & 0.9832          \\
CAR-UNet\cite{28}       & 0.8445          & 0.9850          & 0.9743          & 0.9911 \\
%RBVS-Net\cite{32}                & 0.8188          & 0.9824          & 0.9708          & -               \\
(M)SLAe-Net (Our)       & \textbf{0.8496} & 0.9834          & \textbf{0.9802} & \textbf{0.9925} \\ \hline
\multicolumn{5}{c}{HRF}                                                                         \\ \hline
%Orlando et al.\cite{33} & 0.7874          & 0.9584          & -               & -               \\
Yan et al.\cite{34}     & 0.788           & 0.959           & 0.943           & -               \\
Kamini Upadhyay et al.\cite{35} & 0.750           & \textbf{0.972}  & 0.952           & 0.960           \\ \hline
(M)SLAe-Net ($SA$)   & 0.762           & 0.935           & 0.949           & 0.941           \\
(M)SLAe-Net ($D$-DPP) & 0.789  & 0.939           & 0.958           & 0.962           \\
(M)SLAe-Net (Our)       & \textbf{0.801}  & 0.951           & \textbf{0.961}  & \textbf{0.969}  \\ \hline
\end{tabular}
}
\end{table}

In Table \ref{tab:final_result}, (M)SLAe-Net(SA) depicts that the model consists of SA modules only, (M)SLAe-Net($D$-DPP) means only $D$-DPP modules are present. (M)SLAe-Net(Our) is the final model with all of the modules and components. 

\section{Conclusion and Future Work}
\label{sec:conclusion}

A Multi-Scale, Multi-Level Attention embedded Network ((M)SLAe-Net) for retinal vessel segmentation was proposed in this paper. Our novel dynamic dilated pyramid pooling module and the use of the $SA$ module in our uniquely designed CNN architecture gave a precise, robust, and near complete retinal vessel segmentation. Aggregation of feature maps from all levels is a part of our unique network design and has not been proposed before to the best of our knowledge, virtually giving our model a multi-stage processing design. (M)SLAe-Net outperforms state-of-the-art methods on a notable number of metrics. We firmly believe that our novel $D$-DPP module can be utilised in other standard models for a performance boost, and our network design may be extended to other medical image segmentation tasks.

\section{Acknowledgment}
The authors would like to thank the Department of Computer Science and Engineering, and Department of Electrical Engineering, Indian Institute of Technology Jodhpur, India for providing us with computational resources.  

%\section{Compliance with Ethical Standards}

%\textbf{Conflict of interest} Authors of this paper declare no conflict of interest.
%\textbf{Ethical Approval} This article does not contain any studies with human participants or animals performed by any of the authors. Datasets used in this study were publicly available. 

\bibliographystyle{IEEEbib}
\bibliography{conference_101719}

\end{document}